\documentclass[12pt]{article}
\usepackage[utf8]{inputenc}
\usepackage{amsmath}
\usepackage{amssymb}
\usepackage{graphicx}
\usepackage{lineno}
\usepackage[margin=1.0in]{geometry}
\usepackage[square,comma,sort&compress,numbers]{natbib}
\usepackage{hyperref}
\usepackage{setspace}
\usepackage{enumitem}
\usepackage{verbatim}
\title{A phase field model with arbitrary misorientation dependence of grain boundary energy}

\author{Philip Staublin$^{a,*}$, Yuri Mishin$^{b}$, Peter W. Voorhees$^{c,d,e}$, and James A. Warren$^{f}$} 
\date{\today}

\begin{document}

{\setstretch{1.0}
\maketitle

$^a$Department of Materials Science and Engineering \\University of Michigan \\
2300 Hayward Street \\
Ann Arbor, MI 48109

$^b$Department of Physics and Astronomy \\George Mason University \\
4400 University Drive, MSN 3F3 \\
Fairfax, VA 22030-4444

$^c$Department of Applied Physics and Materials Science \\California Institute of Technology \\
1200 East California Boulevard \\
Pasadena, CA 91125

$^d$Department of Materials Science and Engineering \\Northwestern University \\
2220 Campus Drive  \\
Evanston, IL 60202

$^e$Department of Engineering Sciences and Applied Mathematics \\Northwestern University \\
2220 Campus Drive \\
Evanston, IL 60202

$^f$Material Measurement Laboratory\\ National Institute of Standards and Technology\\ Gaithersburg, MD 20899

$^*$Corresponding author; pstaublin@u.northwestern.edu
}

\linenumbers

\textbf{\large{Abstract}}

Grain growth in polycrystals is often simulated using orientation-field models, which employ a field to represent the local orientation of the crystal lattice. These models can  be challenging to represent a realistic misorientation dependence of grain boundary free energy. We prove that existing orientation-field models, in general, cannot reproduce a decrease in the grain boundary free energy with a increasing misorientation angle, demonstrating a significant limitation of previous models in applications to polycrystalline materials. To overcome this limitation, we propose a modification to the Kobayashi-Warren-Carter model for grain growth wherein the coefficients of the free-energy functional become functions of the misorientation between the grains, which is a non-local quantity. Due to this modification, an arbitrary dependence of the grain boundary free energy on the misorientation can be embedded in the model. We propose calculating the non-local misorientation by interpolating the orientation field at a fixed distance in both directions along the local grain boundary normal vector. The capabilities of the model are demonstrated by introduction of a sharp cusp to the misorientation dependent grain boundary free energy. Finally, we propose an extension of the model to three dimensions.


\modulolinenumbers[5]

\section{Introduction}

Orientation-field models are a class of phase-field models that simulate polycrystalline systems by treating the local crystal orientation as a field variable \cite{KWC1998, KWC2000, KWC2003, HMP2012, HMP2017, Admal2019, Staublin2022, Axel2025preprint}. In contrast to the multi-order-parameter \cite{ChenWang1996, FanChen1997, Moelans2008} and multi-phase-field models \cite{Steinbach1996, Steinbach1999, Nestler1998, GarckeNestler1999, Nestler2005, FolchPlapp2005, Kim2006, SteinbachPFReview, Kim2014}, which assign a distinct field variable to each crystal orientation, the orientation-field models have only a single order parameter coupled to a continuous orientation field. Orientation-field models have the computational advantage of simulating any number of grains without scaling the number of required order parameters; in addition, the triple junction dynamics reproduces the results of classical theory \cite{Staublin2022}. In order to leverage the advantages of the orientation-field models for practical problems, such as the extraction of information from experimental tomographic measurements of grain growth \cite{Zhang2020}, the relationship between model parameters and the properties of grain boundaries in the model must be clearly defined.

There are two families of orientation-field models, differing by the manner in which they ensure localized grain boundaries. The Kobayashi-Warren-Carter model includes two terms dependent on the orientation field $\theta$ in the free energy density: one depends on $|\nabla\theta|$ while the second depends on $|\nabla\theta|^2$. At constant temperature, and without an inclination dependence of the grain boundary (GB) energy, the free energy of the model in a domain $\Omega$ is \cite{KWC2003}
\begin{equation}
\label{eq_kwc}
 F[\phi,\theta] = \int_\Omega \left[V(\phi) + \frac{\alpha^2}{2}|\nabla\phi|^2 +sg(\phi)|\nabla\theta|+\frac{\epsilon^2}{2}h(\phi)|\nabla\theta|^2 \right]dV.
\end{equation}
The coupling functions $g(\phi)$ and $h(\phi)$ increase monotonically, with the influence of orientation vanishing in the liquid phase ($\phi=0$). The potential $V(\phi)$ is a double well potential in a solid-liquid system with minima in the two phases.  For a grain-only system, $V(\phi)$ only has a minimum in the solid phase. The model parameters $\alpha$, $s$, and $\epsilon$ determine the GB energy\footnote{. Generally speaking, the model predicts the grain boundary \emph{free} energy, but we will refer to it as grain boundary energy for brevity.} by adjusting the strength of the coupling between $\phi$ and $\theta$. The choice of model parameters affects a critical misorientation at which the GB width diverges, effectively resulting in melting of the boundary. The presence of the first order term results in stationary solutions for the GB profile.

The second family of models, the HMP model after Henry, Mellenthin, and Plapp \cite{HMP2012}, omits the first-order gradient term in $\theta$, favoring a singular coupling function on the $|\nabla\theta|^2$ term to localize the boundaries. The free energy in the HMP model is defined by
\begin{equation}
\label{eq_hmp}
F[\phi,\theta] = \int_\Omega\left[\frac{W_\phi^2}{2}|\nabla\phi|^2+HV(\phi)+\frac{W_\theta^2}{2}g(\phi)|\nabla\theta|^2\right]dV,
\end{equation}
where $W_\phi$ and $W_\theta$ are gradient energy coefficients, $H$ scales the single-well potential $V(\phi)$, and $g(\phi)$ is a coupling function that diverges at $\phi=1$. This divergence plays two roles: firstly localizing the gradient in orientation to a finite width GB, and secondly, preventing changes to the orientation field outside of the GB region. The latter amounts to removing grain rotations from the model, by creating an infinite driving force against rotation in the bulk crystal where $\phi=1$. This model was adopted to simulate grain growth using a single-well potential $V(\phi)$, and properties such as triple junction angles and steady-state velocities were verified using simulations \cite{Staublin2022}.

Both the KWC and HMP models contain an explicit dependence of the GB energy on the gradient in orientation, preventing fully arbitrary misorientation dependence of the boundary energy without further modification. Previous work has demonstrated methods for introducing the dependence of the GB energy on the inclination and misorientation axis \cite{Admal2019}. However, the GB energy still increases with the misorientation angle for a given inclination and misorientation axis. As will be shown in the following, this is a fundamental limitation, which prevents the modeling of both isotropic GB energies (the soap froth problem) and, more importantly, realistic polycrystalline materials, where there are often multiple minima in the boundary energy. 

In this work, we propose an alternative formulation for the KWC model, motivated by the preceding analysis. As we show analytically in Section \ref{sec_theory}, the GB energy must increase monotonically with misorientation in the existing formulations of both models  (KWC and HPM). The proposed reformulation of the KWC model introduces a dependence of the model coefficients on a nonlocal measure of GB misorientation. This choice is inspired by the work of Han and van de Walle \cite{Axel2025preprint}, who formulate a new orientation-field model and an iterative algorithm for computing misorientation non-locally. In Section \ref{sec_methods}, we provide the modified KWC model equations and an algorithm to rapidly compute nonlocal misorientation. We discuss the introduction of misorientation dependence of the GB energy, as well as adjustment of the orientation-field mobility to prevent grain rotation (``plateau motion''). In Section \ref{sec_results}, we examine the energy and mobility of GBs in the model and demonstrate the ability of the model to capture sharp cusps in GB energy. Finally, we describe a 3D extension to the model in Section \ref{sec_3D} and present our conclusions in Section \ref{sec_conclusions}.

\section{Limitations of existing models}
\label{sec_theory}
\subsection{Equilibrium Bicrystal Solutions}

The goal of this section is to provide an analytical result for the GB energy as a function of misorientation for the case of a single planar GB located at the center of an infinite one-dimensional domain. The equations derived in this section will be used in section \ref{Dependence_on_delta_theta} to show that, within the existing models, the GB energy must increase with the misorientation angle. 

The bulk crystal orientations are $\theta_1$ and $\theta_2$ in the left and right halves of the domain, respectively. The order parameter $\phi$ is $\phi=1$ in the bulk crystal and decreases near the boundary. We assume that the order parameter profile $\phi(x)$ has a single minimum in the GB region, while the misorientation profile $\theta(x)$ is a monotonic function of the coordinate $x$. The boundary conditions are summarized as follows:
\begin{equation}
\label{eq2_bc_theta1}
    \lim_{x \to -\infty}\theta(x)=\theta_1,
\end{equation}
\begin{equation}
\label{eq2_bc_theta2}
    \lim_{x \to +\infty}\theta(x)=\theta_2,
\end{equation}
\begin{equation}
\label{eq2_bc_phi_infty}
    \lim_{x \to \pm\infty}\phi(x)=1.
\end{equation}

In addition to these boundary conditions, the potential is zero in the bulk crystal, $V(1)=0$, so the only contribution to the free energy is due to the presence of the planar GB. The free energy contribution from the GB is positive, so $V(\phi)>0$; these are the only constraints imposed on the potential. No assumptions are made about the form of $g(\phi)$ and $h(\phi)$.

To determine the change in GB energy with misorientation in the KWC model, we begin by finding the equilibrium profiles and corresponding free energy for an arbitrary misorientation that is set by the boundary conditions. Much of this derivation can be found in Appendix A of Warren et al. \cite{KWC2003}. We adopt the convention of $x$-subscripts for the spatial derivatives of $\phi$ and $\theta$, i.e. $\phi_x\equiv d\phi/dx$, $\phi_{xx}\equiv d^2\phi/dx^2$, and $\theta_x\equiv d\theta/dx$. We retain $d/dx$ for more general differentiation with respect to $x$. The Euler-Lagrange equations for the KWC model in one dimension are:
\begin{equation}
\label{eq_kwc_el_phi}
    0 = \frac{\delta F}{\delta\phi} = \frac{ dV}{d\phi}
    + s\frac{dg}{d\phi}\left|\theta_x\right|
    + \frac{\epsilon^2}{2}\frac{dh}{d\phi}\left(\theta_x\right)^2
    - \alpha^2 \phi_{xx}
\end{equation}
and
\begin{equation}
\label{eq_kwc_el_theta}
    0 = \frac{\delta F}{\delta\theta} = \frac{d}{dx}\left[sg(\phi)\theta_x\left|\theta_x\right|^{-1} + \epsilon^2 h(\phi)\theta_x\right].
\end{equation}

The first integral, a direct consequence of Noether's theorem \cite{gelfand_fomin}, of the system is found to be
\begin{equation}
    V(\phi)-\frac{\alpha^2}{2}\left(\phi_x\right)^2-\frac{\epsilon^2}{2}h(\phi)\left(\theta_x\right)^2=0.
\label{FI}
\end{equation}

The $\theta$ equation, Eq. (\ref{eq_kwc_el_theta}), can be trivially integrated once, introducing a constant of integration $C$:
\begin{equation}
    C = sg(\phi)\theta_x\left|\theta_x\right|^{-1} + \epsilon^2 h(\phi)\theta_x.
\end{equation}
This equation can be written as
\begin{equation}
    C =\pm sg(\phi) + \epsilon^2 h(\phi)\theta_x, \label{C}
\end{equation}
where we use the positive sign if $\Delta \theta \equiv \theta_2 - \theta_1>0$ and the negative if $\Delta \theta <0$.
Solving for $\theta_x$, we find
\begin{equation}
\label{eq_dthetadx}
    \theta_x = \frac{C \mp sg(\phi)}{\epsilon^2 h(\phi)}.
\end{equation}

Equations (\ref{FI}) and (\ref{eq_dthetadx}) show that the profiles must possess the following symmetries: $\phi(x)=\phi(2\zeta_0-x)$ and $\theta_x(x)=\theta_x(2\zeta_0-x)$, where $\zeta_0$ is the coordinate of the minimum of $\phi(x)$. Indeed, Eq.(\ref{eq_dthetadx}) shows that at points with the same $\phi$ on either side of the minimum, $\theta_x$ must be the same, while according to Eq.(\ref{FI}), at such points the derivatives $\phi_x$ have the same magnitude with opposite signs.

It has been shown  \cite{KWC2003} that the misorientation profile $\theta(x)$ is localized in a finite interval $\zeta_1\leq x \leq \zeta_2$ that covers the GB core with $\theta_x \equiv 0$ outside this interval. In contrast, the order-parameter profile $\phi(x)$ has diffuse tails that extend outside the interval of nonzero $|\theta_x|$ values. Consequently, solutions are sought in three regions: the inner region $[\zeta_1,\zeta_2]$ and two outer regions $[-\infty,\zeta_1]$   and $[\zeta_2,\infty]$. By the symmetry of the profiles,
\begin{equation}
    \phi(\zeta_1)=\phi(\zeta_2) \equiv\phi_{max}
    \label{phi_max}
\end{equation}

Since at the end points of the inner region  $\theta_x=0$ and $\phi = \phi_{max}$, we have from Eq.(\ref{eq_dthetadx}):
\begin{equation}
    C = \pm sg(\phi_{max}).
\end{equation}
This allows us to rewrite Eq.(\ref{C}) as 
\begin{equation}
        \theta_x = \pm \frac{sg(\phi_{max}) - sg(\phi)}{\epsilon^2 h(\phi)}.
        \label{dt/dx}
\end{equation}
In the outer region, because $\theta_x=0$, 
\begin{equation}
\label{eq_kwc_dphidx_outer}
    \frac{\alpha^2}{2}\left(\phi_x\right)^2 = V(\phi).
\end{equation}
giving a direct relationship between the spatial variation in $\phi$ and the potential.  
However, to obtain $\phi_x$ in the inner region, we substitute Eq. (\ref{dt/dx}) into Eq. (\ref{FI}) and find
\begin{equation}
\label{eq_kwc_dphidx}
    \frac{\alpha^2}{2}\left(\phi_x\right)^2 = V(\phi)
    - \frac{s^2}{2\epsilon^2}\frac{\left(g(\phi_{max})-g(\phi)\right)^2}{h(\phi)}.
\end{equation}
Applying this equation to the minimum of the order parameter, we have  
\begin{equation}
\label{phi_0}
    V(\phi_0) =
     \frac{s^2}{2\epsilon^2}\frac{\left(g(\phi_{max})-g(\phi_0)\right)^2}{h(\phi_0)}.
\end{equation}
On the other hand, we integrate Eq.(\ref{dt/dx}) through the inner region:
\begin{equation}
    \int_{\zeta_1}^{\zeta_2} \theta_x dx= \Delta\theta
    = \pm \int_{\zeta_1}^{\zeta_2}\left[\frac{sg(\phi_{max}) - sg(\phi)}{\epsilon^2 h(\phi)}\right]dx
    = \pm \int_{\zeta_1}^{\zeta_2}\left[\frac{sg(\phi_{max}) - sg(\phi)}{\epsilon^2 h(\phi)}\right] \frac{dx}{d\phi} d\phi
    \label{delta_theta}.
\end{equation}
Inserting $\phi_x$ from Eq.(\ref{eq_kwc_dphidx_outer}), we have
\begin{equation}
    \Delta\theta = \pm \frac{\sqrt{2}\alpha s}{\epsilon^2}\int_{\phi_0}^{\phi_{max}}\frac{g(\phi_{max}) - g(\phi)}{\epsilon^2 h(\phi) \displaystyle \sqrt{V(\phi)-\frac{s^2}{2\epsilon^2}\frac{[ g(\phi_{max}) - sg(\phi)]^2}{h(\phi)}}}d\phi
    \label{Delta_t}
\end{equation}

Equations (\ref{phi_0}) and (\ref{Delta_t}) establish a relationship between the minimum/maximum order parameters $\phi_0$ and $\phi_{max}$ and $\Delta\theta$  \cite{KWC2003}. Knowing $\phi_0$ and $\phi_{max}$, the boundaries of the inner region can be found from \cite{KWC2003}
\begin{equation}
    \zeta_2= \zeta_0 + \displaystyle \frac{\alpha}{\sqrt{2}}   \int_{\phi_0}^{\phi_{max}}\frac{d\phi} {\displaystyle \sqrt{V(\phi)-\frac{s^2}{2\epsilon^2}\frac{[ g(\phi_{max}) - sg(\phi)]^2}{h(\phi)}}}
    \label{zeta}
\end{equation}
and $\zeta_1 = 2\zeta_0-\zeta_2$.

\subsection{Dependence of grain boundary energy on $\Delta\theta$}\label{Dependence_on_delta_theta}

In this section, we derive an analytical result for the derivative of $\gamma_{GB}$ with respect to the misorientation angle $\Delta\theta$. The result is obtained by calculating the variational derivative of the free-energy functional with respect to $\Delta\theta$ and is general enough to be applicable to both the KWC and HMP models, as well as similar phase-field functionals.

We assume that the system is configured as detailed above, with the GB centered at $x=\zeta_0$. The analysis holds for the KWC and HMP models. We define a function $w$ to represent the argument of Eq. (\ref{eq_kwc}), writing
\begin{equation}
\label{gamma-gb-x}
\gamma_{GB}=\int_{\zeta_1}^{\zeta_2}w( \theta_{x}(x),\phi(x),\phi_{x}(x))dx +\int_{-\infty}^{\zeta_1} w(\phi(x),\phi_x(x)) dx +\int_{\zeta_2}^\infty w(\phi(x),\phi_x(x)) dx
\end{equation}
with the following boundary conditions: $\theta_x=0$ at $x \leq \zeta_1$ and $x \geq \zeta_2$, and $\theta(\zeta_2)-\theta(\zeta_1)=\Delta\theta$. Note that we have omitted $\theta_x$ from the arguments of $w$ outside the range $\zeta_1\leq x \leq \zeta_2$ where $\theta_x$ is 0. 

We revisit the variational derivative formally, now retaining the boundary terms, as this is precisely what we are interested in varying (specifically $\Delta\theta$).  These give natural boundary conditions for the first variation in the GB free energy that specify the values of $\theta$ at each end of the domain. This is critical in setting the dependence of the GB energy on misorientation. The first variation of the GB energy is
\begin{eqnarray}
 \delta\gamma_{GB} &=& \int_{\zeta_1}^{\zeta_2} \left[
        \left(- \frac{d}{dx}\frac{\partial w}{\partial \theta_x}\right)\delta\theta
      + \left(\frac{\partial w}{\partial \phi}   - \frac{d}{dx}\frac{\partial w}{\partial \phi_x}  \right)\delta\phi
    \right]dx \nonumber\\
     &+& \int_{-\infty}^{\zeta_1} \left[
        \left(\frac{\partial w}{\partial \phi} - \frac{d}{dx}\frac{\partial w}{\partial \phi_x}\right)\delta\phi
    \right]dx + \int_{\zeta_2}^{\infty} \left[
        \left(\frac{\partial w}{\partial \phi} - \frac{d}{dx}\frac{\partial w}{\partial \phi_x}\right)\delta\phi
    \right]dx\nonumber\\
    &+&   \frac{dw}{d\theta_x}\,\delta\theta\,\Bigg|_{\zeta_1}^{\zeta_2}
    + \frac{dw}{d\phi_x}\,\delta\phi\,\Bigg|_{\zeta_1}^{\zeta_2}
    + \frac{dw}{d\phi_x}\,\delta\phi\,\Bigg|_{-\infty}^{\zeta_1}
    + \frac{dw}{d\phi_x}\,\delta\phi\,\Bigg|_{\zeta_2}^{\infty}\nonumber\\
    &+&w( \theta_{x}(\zeta_2),\phi(\zeta_2),\phi_{x}(\zeta_2))\delta\zeta_2 - w( \theta_{x}(\zeta_1),\phi(\zeta_1),\phi_{x}(\zeta_1))\delta\zeta_1\nonumber\\
    &+& w(\phi(\zeta_1),\phi_{x}(\zeta_1))\delta\zeta_1 -w(\phi(\zeta_2),\phi_{x}(\zeta_2))\delta\zeta_2.
    \label{eq:1}
\end{eqnarray}
Note that we included variations of the boundaries $\zeta_1$ and $\zeta_2$ of the three regions discussed in the previous section. This was necessary because these boundaries are related to the angles of misorientation through equations (\ref{phi_0}) and (\ref{Delta_t})-(\ref{zeta}).  The integrals in Eq. (\ref{eq:1}) are, of course, zero. Also, 
$dw/d\phi_x=\alpha^2\phi_x \rightarrow 0$ at $x \rightarrow \pm \infty$. To preserve the symmetry of the order parameter profile, the variations $\delta\phi$ at $\zeta_1$ and $\zeta_2$ must be equal. As a result, the terms with $\delta\phi$ cancel mutually.

In the last two lines of Eq.(\ref{eq:1}), the pairs of terms with $\delta\zeta_1$ and $\delta\zeta_2$ mutually cancel because $\theta_x=0$ at $\zeta_1$ and $\zeta_2$, and therefore 
\begin{equation}
    w( \theta_{x}(\zeta_1),\phi(\zeta_1),\phi_{x}(\zeta_1))=w( \phi(\zeta_1),\phi_{x}(\zeta_1)),
\end{equation}
\begin{equation}
    w( \theta_{x}(\zeta_2),\phi(\zeta_2),\phi_{x}(\zeta_2))=w( \phi(\zeta_2),\phi_{x}(\zeta_2)).
\end{equation}

We are left with the following equation:
\begin{equation}
    \delta\gamma_{GB} =\frac{dw}{d\theta_x}\,\delta\theta\,\Bigg|_{\zeta_1}^{\zeta_2}.
\end{equation}
We are reminded from Eq. (\ref{eq_kwc_el_theta}) that $\partial w/\partial\theta_{x}=C$, and therefore
\begin{equation}
    \delta\gamma_{GB} =C\delta\theta\,\Bigg|_{\zeta_1}^{\zeta_2}=C[\delta\theta_{x={\zeta_2}}-\delta\theta_{x={\zeta_1}}].
\end{equation}  
Importantly for this analysis, $\delta\theta_{x={\zeta_2}}-\delta\theta_{x={\zeta_1}}=\delta(\Delta\theta)$, from which
\begin{equation}
    \delta\gamma_{GB} =C\delta (\Delta \theta).
\end{equation}
In the KWC model, $C=\pm sg(\phi_{max})$.  It would be a different constant for the HMP model, but the variation of the energy with respect to $\theta$ must be constant because it lacks a dependence on $\theta$, only $\theta_x$. Thus, for the KWC model we have 
\begin{equation}
\delta \gamma_{GB}=\pm sg(\phi_{max})\delta(\Delta\theta).\label{eq:3}
\end{equation}
In other words,
\begin{equation}
\dfrac{{d}\gamma_{GB}}{{d}|\Delta\theta|}=sg(\phi_{max}).\label{eq_kwc_dgamma_final}
\end{equation}

Equation (\ref{eq_kwc_dgamma_final}) shows that the GB energy must remain constant or increase with misorientation in the KWC model. In fact, as the above approach is quite general, particularly from Eq. (\ref{eq:1}), it can be generalized to other orientation-field models. This result precludes the application of orientation-field models to isotropic systems (i.e., one cannot generally simulate the ``soap-froth'' problem with an orientation-field model). In addition, cusps in the GB energy at special misorientations cannot be reproduced without further modification of the model's free energy functional. On the basis of these conclusions, polycrystalline systems with an arbitrary five-dimensional dependence of the GB energy on GB crystallography cannot be modeled using such, as formulated, orientation-field models.

\section{Reformulation of the KWC model}
\label{sec_methods}

In this section, we propose a generalized formulation of the KWC model that addresses the fundamental inability of existing models to predict any realistic misorientation dependence of the GB energy.

We considered a number of methods by which the limitation imposed by Eq. (\ref{eq_kwc_dgamma_final}) could be circumvented or eliminated, including introducing additional order parameters, allowing $V(\phi)$, $g(\phi)$, and/or $h(\phi)$ to take negative values, and introducing higher-order terms in $\nabla\theta$. However, each of these methods introduces new problems, such as limiting the range of $\Delta\theta$ for which equilibrium solutions exist or dramatically increasing the mathematical and computational complexity. Instead, we argue that the root of the problem is the non-locality of information required to determine GB misorientation. 

If a GB is defined as a diffuse region over which $\theta$ changes from one constant value to another, the misorientation can be computed by integrating $\nabla\theta$ along the GB normal direction, $\vec{n}$, through the entire diffuse boundary region, $-d<l<+d$, $\nabla\theta(\pm d)=0$:
\begin{equation}
\label{eq_nonlocal_dtheta}
    \Delta\theta = \int_{-d}^{+d}\nabla\theta \cdot\vec{n} dl
    = \theta(\vec{x}+d\vec{n}) - \theta(\vec{x}-d\vec{n}).
\end{equation}
If, as is often assumed in phase-field models, the diffuse interface width is greater than the discrete elements on which the solution is represented ($d\gg\Delta x$), then the misorientation is a nonlocal quantity compared to the local energy density typically considered in phase-field models. Although there is a relationship between $\nabla\theta$ and $\Delta\theta$, $\Delta\theta$ cannot be calculated from $\nabla\theta$ at only a single point where the equations are numerically integrated. However, based on the (potentially incomplete) prevailing theory of grain growth, $\Delta\theta$ should determine the GB energy rather than $\nabla\theta$. 

To reconcile this, we propose introducing a dependence on the nonlocal misorientation to the KWC model's coupling coefficients, $s$ and $\epsilon$. This idea is inspired by the recent model of Han and van de Walle \cite{Axel2025preprint}, in which the nonlocal misorientation was introduced and an iterative search algorithm was proposed to calculate it. In the following section, we introduce the nonlocal misorientation to the KWC model and describe an  algorithm to calculate $\Delta\theta$.

The free energy functional of the KWC model (Eq. (\ref{eq_kwc})) is modified by introducing a dependence on the nonlocal misorientation, $\Delta\theta$, as follows:
\begin{equation}
\label{eq_nonlocal_kwc}
     F[\phi,\theta] = \int_\Omega \left[V(\phi) + \frac{\alpha^2}{2}|\nabla\phi|^2 +s(\Delta\theta)g(\phi)|\nabla\theta|+\frac{s(\Delta\theta)^2\epsilon^2}{2}h(\phi)|\nabla\theta|^2 \right]dV.
\end{equation}
Note that $\Delta\theta$ represents the nonlocal misorientation, not the Laplacian operating on $\theta$. The choice to introduce a dependence of $s$ on $\Delta\theta$ is motivated by Eq. (\ref{eq_kwc_dgamma_final}): if $g(\phi)$ must increase monotonically, then $s$ must be allowed to vary so that $\gamma_{GB}$ can decrease or increase with misorientation. 

The Euler-Lagrange equation for $\theta$ now involves variations in $\Delta\theta$; however, assuming that $d$ is large enough to satisfy the condition $\nabla\theta(\vec{x}\pm d\vec{n})=0$, the variation in nonlocal misorientation is zero to first order \cite{Axel2025preprint}. 
Then, the evolution equations closely resemble those of the KWC model:
\begin{equation}
\label{eq_phi_evolve}
    \frac{\partial\phi}{\partial t} = -M_{\phi}\frac{\delta F}{\delta\phi}
    = -M_{\phi}\left[V'(\phi)+s(\Delta\theta)g'(\phi)|\nabla\theta|
    + \frac{s(\Delta\theta)^2\epsilon^2}{2}h'(\phi)|\nabla\theta|^2
    - \alpha^2 \nabla^2\phi\right]
\end{equation}
and
\begin{equation}
\label{eq_theta_evolve}
    \frac{\partial\theta}{\partial t} = -M_{\theta}\frac{\delta F}{\delta\theta}
    = M_{\theta}\left[\nabla\cdot\left(s(\Delta\theta)^2\epsilon^2 h(\phi)\nabla\theta + s(\Delta\theta)g(\phi)\frac{\nabla\theta}{|\nabla\theta|}\right)\right].
\end{equation}

\section{Computational methodology}

\subsection{Numerical implementation of the model}

The proposed model differs from the previous models in that the evolution equations for the misorientation and disorder parameter fields depend on the global misorientation angle $\nabla\theta$. Therefore, the new and critical part of the numerical implementation of the model is the calculation of the global parameter $\nabla\theta$ based on local fields.

To compute the nonlocal misorientation at a given point, two quantities must be determined: the GB normal vector, $\vec{n}$, and ``search distance'', $d$, which enters the bounds of the integral in Eq. (\ref{eq_nonlocal_dtheta}). The former can be determined based on $\nabla\theta$. The latter is more complex, as the search distance should result in $\nabla \theta(\vec{x}\pm d\vec{n})=0$ for the point $\vec{x}$ at which we are evaluating $\Delta\theta$. Han and van de Walle \cite{Axel2025preprint} use an iterative search algorithm, wherein $d$ depends on the $\theta$ field and is chosen by incrementally advancing along $\vec{n}$ until $\nabla\theta$ is less than a set ``cutoff'' value. For bicrystals, this approach ensures that $\nabla\theta(\vec{x}\pm d\vec{n})=0$ is always satisfied. Within a triple junction, if $\vec{n}$ points along a planar GB, the iterative search algorithm will not terminate as a region where $\nabla\theta(\vec{x}+d\vec{n})=0$ will never be reached. To avoid this problem, Han and van de Walle set a maximum search distance for their algorithm.

The computational cost to compute $\Delta\theta$ can be significantly reduced by introducing a constant search distance $d$. Treating $d$ as a constant eliminates the iterative process and reduces the number of 2D interpolations to the minimum number of 2 per evaluation of $\Delta\theta$. However, $d$ must be chosen to be large enough to cover the entire GB width while small enough to avoid intersecting additional grain boundaries. For example, with $d$ much larger than the GB width and considering a circular grain embedded within a matrix, the nonlocal misorientation could be incorrectly evaluated as zero if $d$ is larger than the radius of the embedded grain.

We use the finite volume method to discretize Eqs.~(\ref{eq_phi_evolve}) and (\ref{eq_theta_evolve}) in space, and perform integration with respect to time by the method of lines using an explicit low-storage fourth-order Runge-Kutta scheme with an embedded fifth-order scheme for adaptive time stepping \cite{Williamson1980, Carpenter1994}. The model is implemented in a serial C++ code, which is sufficient to produce the results presented here. Parallelization of the nonlocal misorientation algorithm is challenging, as each processor requires long-range information about the $\theta$ field. The usual cartesian topology for an MPI implementation would require a layer of ghost points covering a region of at least $d$ in size; alternatively, the entire $\theta$ field would need to be shared or broadcast among every processor. Parallel implementation of the model code is left for future work.

A final consideration must be made to address the periodic nature of $\theta$. We constrain values of theta to the interval $(-\pi/2,\pi/2]$, adding or subtracting $\pi$ at each timestep to ensure $\theta$ remains in this range. Differences between angles must be defined such that this symmetry is captured. We implement an angle difference operator which adds or subtracts a correction factor of $\pi$ \cite{KWC2003}:
\begin{equation}
    \text{Diff}(\theta_1,\theta_0) =
    \begin{cases}
        \theta_1 - \theta_0 + \pi, & \theta_1-\theta_0 < -\pi/2 \\
        \theta_1 - \theta_0,       & -\pi/2 <\theta_1 - \theta_0\leq \pi/2 \\
        \theta_1 - \theta_0 - \pi, & \theta_1 - \theta_0 \geq \pi/2
    \end{cases}
\end{equation}
This difference operator is used to compute the gradients in $\theta$ at the voxel faces and centers.

\subsection{Parameterization of the model}

We adopt the following forms for $V(\phi)$, $g(\phi)$, and $h(\phi)$ from previous work \cite{KWC2003}:
\begin{equation}
    V(\phi) = \frac{a^2}{2}(1-\phi)^2
\end{equation}
and
\begin{equation}
    g(\phi) = h(\phi) = \phi^2.
\end{equation}

Importantly, the functional form of $s(\Delta\theta)$ can be determined from the expression of the GB energy as a function of misorientation. In the limit where $\epsilon=0$, Eq. (\ref{gamma-gb-x}) gives the GB energy as a function of $\phi_0$ and $\Delta\theta$:
\begin{equation}
\label{eq_kwc_gamma_param}
    \gamma_{GB} = sg(\phi_0)\Delta\theta + 2\alpha\int_{\phi_0}^1\sqrt{2V(\phi)}d\phi
    = s\phi_0^2\Delta\theta - 2a\alpha(1-\phi_0).
\end{equation}
When $\epsilon=0$, Eq. (\ref{eq_kwc_el_phi}) can be integrated to obtain $\phi_0$ \cite{KWC2000}:
\begin{equation}
\label{eq_kwc_phi0}
    \phi_0 = \frac{1}{1 + s\Delta\theta/a\alpha}.
\end{equation}
Substituting Eq. (\ref{eq_kwc_phi0}) into Eq. (\ref{eq_kwc_gamma_param}) gives:
\begin{equation}
    \gamma_{GB} = \frac{s\Delta\theta}{1 + s\Delta\theta/a\alpha}.
\end{equation}
With some algebra, we obtain:
\begin{equation}
\label{eq_kwc_s_deltatheta}
    s(\Delta\theta) = \frac{\gamma_{GB}}{(1 - \gamma_{GB}/a\alpha)\Delta\theta}.
\end{equation}
Using Eq. (\ref{eq_kwc_s_deltatheta}), we can embed arbitrary relationships between GB energy and misorientation, allowing the model to naturally capture features such as cusps and symmetry.

\section{Simulation Results}
\label{sec_results}

\subsection{Properties of planar grain boundaries}\label{planar_GB}

\begin{figure}
    \centering
    \includegraphics[width=0.9\linewidth]{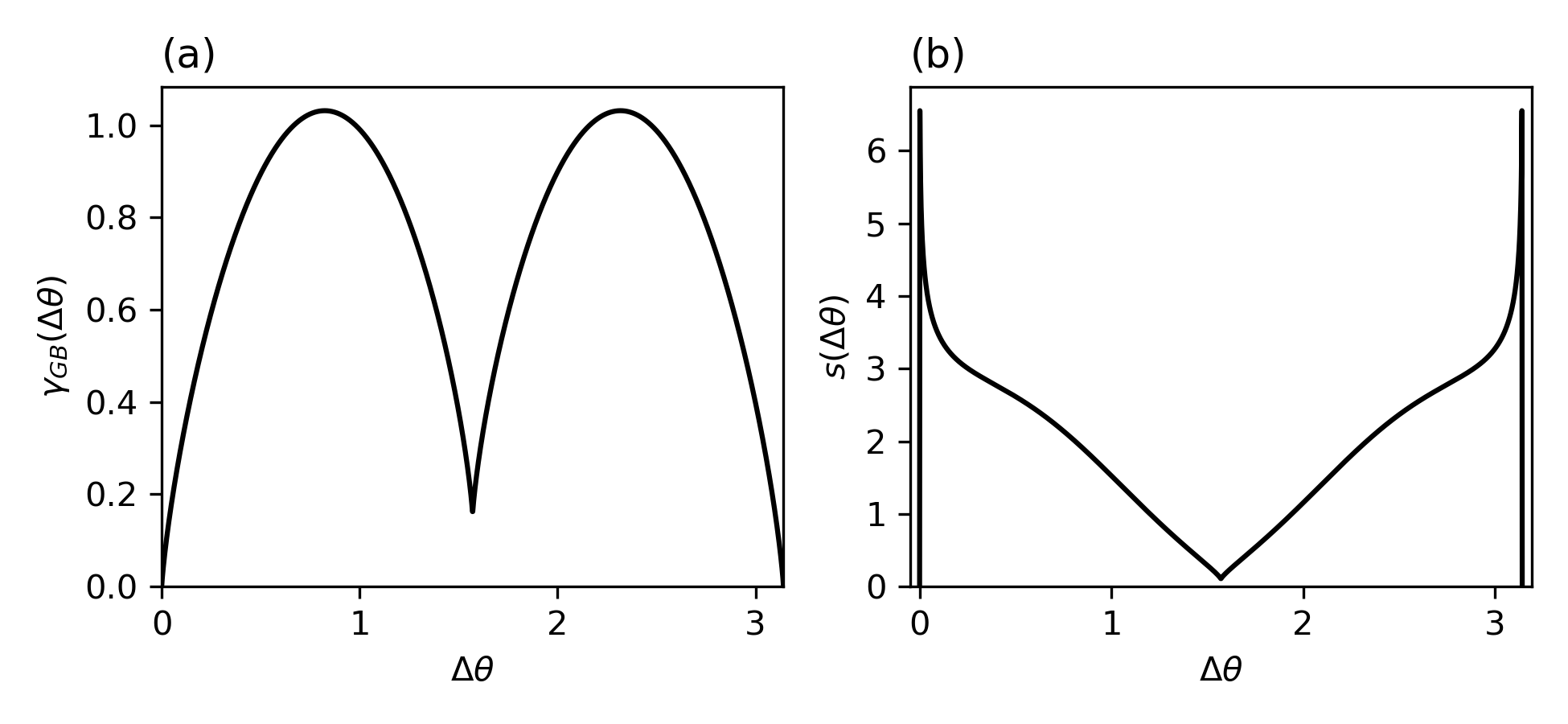}
    \caption{(a) Grain boundary energy function including a cusp at $\Delta\theta=\pi/2$ used in the present results. (b) Coefficient of the linear $\nabla\theta$ term as a function of nonlocal misorientation, $s(\Delta\theta)$, which results from Eq. \ref{eq_kwc_s_deltatheta} given the grain boundary energy function shown in (a).}
    \label{figure_s_gamma}
\end{figure}

To demonstrate the capability of the model, we first apply it to a 1D case representing a planar GB is a bicrystal. We adopt the following expression for $\gamma_{GB}$ in Eq. (\ref{eq_kwc_s_deltatheta}):
\begin{equation}
\label{eq_simulation_gamma}
    \gamma_{GB}(\Delta\theta)=
    \begin{cases}
        0.1|\Delta\theta| + RS(|\Delta\theta|)\cdot RS(\frac{\pi}{2}-|\Delta\theta|)
        & 0 \leq |\Delta\theta| < \frac{\pi}{2} \\
        0.1|\Delta\theta|
        & \Delta\theta = \frac{\pi}{2} \\
    \end{cases}
\end{equation}
where $RS(x)$ is the following function that resembles the Read-Shockley dependence of GB energy on misorientation for low angle GBs:
\begin{equation}
    RS(x) = x(1 - \ln(x)).
\end{equation}
Eq. \ref{eq_simulation_gamma} has a cusp at $\Delta\theta = \pi/2$ and is symmetric on $[0,\pi)$. When $\Delta\theta <0$ or $\Delta\theta \geq \pi$, $\pi$ is added to or subtracted from $\Delta\theta$ until $\Delta\theta\in[0,\pi)$. For the coefficient of the potential, we choose $a=\sqrt{2}$, and for the gradient energy coefficient in $\phi$, we choose $\alpha=2$. Figure \ref{figure_s_gamma} shows the value of $\gamma(\Delta\theta)$ and $s(\Delta\theta)$ using these parameters. Note that $s(\Delta\theta)$ is singular at $\Delta\theta=0$; to avoid numerical problems, we add a small value ($10^{-12}$) to the denominator in Eq. (\ref{eq_kwc_s_deltatheta}). The coefficient of the term $(\nabla\theta)^2$ is $\epsilon=0.5$ except where otherwise noted.

As in previous iterations of the KWC model \cite{KWC2003}, the $\theta$ mobility is divided by a function that is $1$ above a small finite $|\nabla\theta|$ but tends to infinity as $|\nabla\theta|$ approaches zero:
\begin{equation}
\label{eq_theta_mob}
    M_\theta^{-1}=P(\epsilon |\nabla\theta|)=1 - \exp(-\beta\epsilon|\nabla\theta|)
    + \frac{\mu}{\epsilon}\exp(-\beta\epsilon|\nabla\theta|),
\end{equation}
where $\beta$ and $\mu$ are constants that determine the ``cut-off'' in $|\nabla\theta|$. Modifying the mobility in this manner mitigates the so-called ``elevator motion'' where the orientation of bulk grains changes far from the interface. Such a bulk rotation could not be accommodated in a real crystalline material and is therefore unphysical. For the constants in Eq. (\ref{eq_theta_mob}) we choose $\beta=2000$ and $\mu=10^4$. The grid spacing is $\Delta x=0.1$, and the search distance for the nonlocal misorientation is $d=3$ (30 voxels). Zero-flux boundary conditions are used for both $\phi$ and $\theta$, and if the nonlocal misorientation search would extend beyond the simulation domain, the search would end at the point on the domain boundary intersected by the GB normal vector.

\begin{figure}
    \centering
    \includegraphics[width=0.9\linewidth]{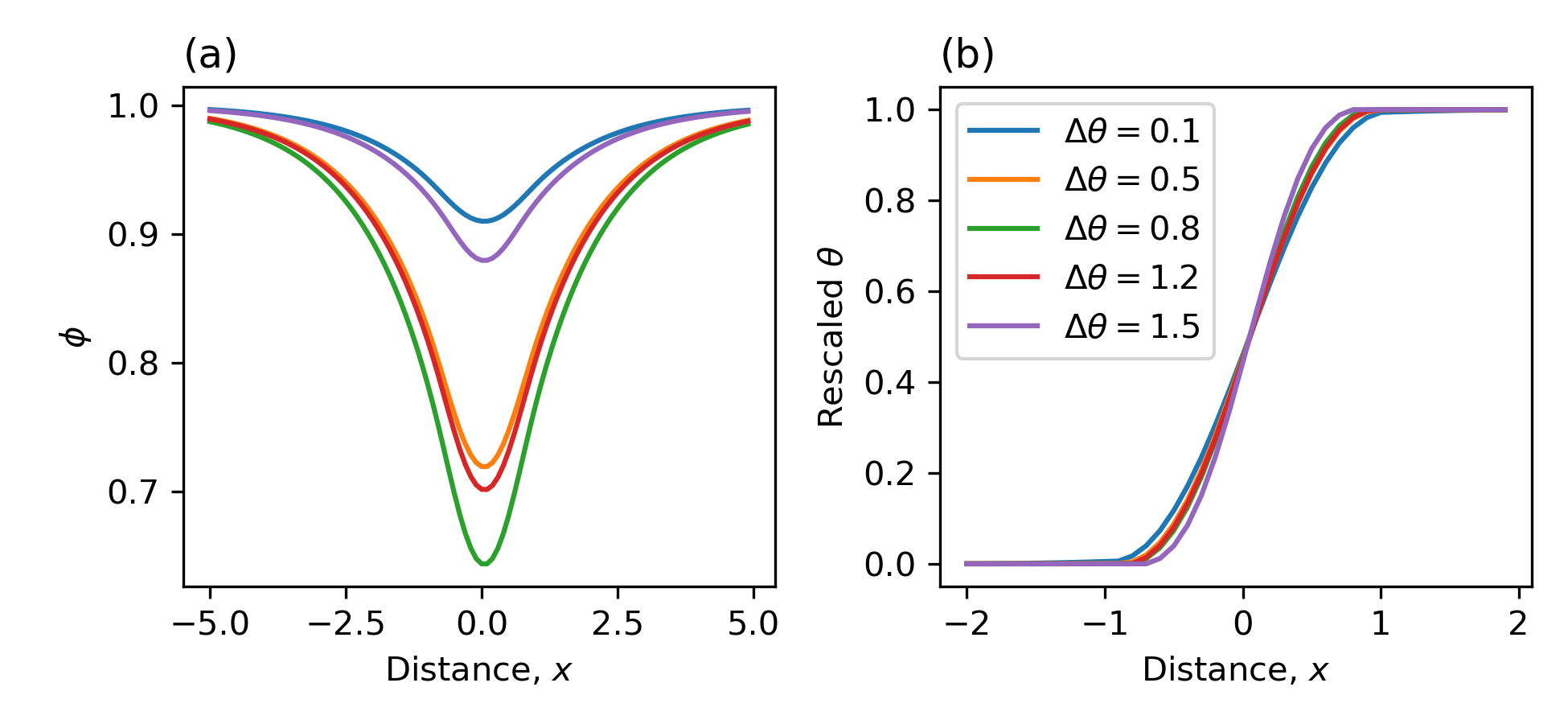}
    \caption{Equilibrium profiles for varying misorientation with $s(\Delta\theta)$ given by Eq. (\ref{eq_kwc_s_deltatheta}) with $\epsilon=0.5$, for (a) $\phi$ and (b) $\theta$. The colors described in the legend on (b) apply to both subfigures.}
    \label{figure_eq_profiles}
\end{figure}

Equilibrium 1D profiles as a function of misorientation are determined by starting from constant $\phi$ and an initially sharp $\theta$ with the jump centered in the computational domain. The equilibrium profiles in $\phi$ have a minimum centered at the boundary, with the minimum value, $\phi_0$, increasing with $\gamma_{GB}$ rather than with misorientation (Figure \ref{figure_eq_profiles}a). The $\theta$ profiles adopt the expected hyperbolic-tangent-like shape; however, it is important to note that the GBs have a truly finite width, where $\nabla\theta$ completely vanishes outside the diffuse interface region, rather than approaching zero in the limit towards infinity. The finite GB width varies with GB energy, increasing slightly as $\gamma_{GB}$ decreases (Figure \ref{figure_eq_profiles}b). Because the dependence on $\Delta\theta$ was introduced to both $\nabla\theta$-dependent terms in the free energy functional, the variation in GB width with GB energy is small (see Eq.~(A.13) in \cite{KWC2003}). This ensures that a single constant $d$ can be chosen for the calculation of nonlocal misorientation. If the GB width were to vary more significantly, the nonlocal misorientation would need to be determined by an alternative algorithm, such as the iterative search of \cite{Axel2025preprint}. 

\begin{figure}[hbt]
    \centering
    \includegraphics[width=0.5\linewidth]{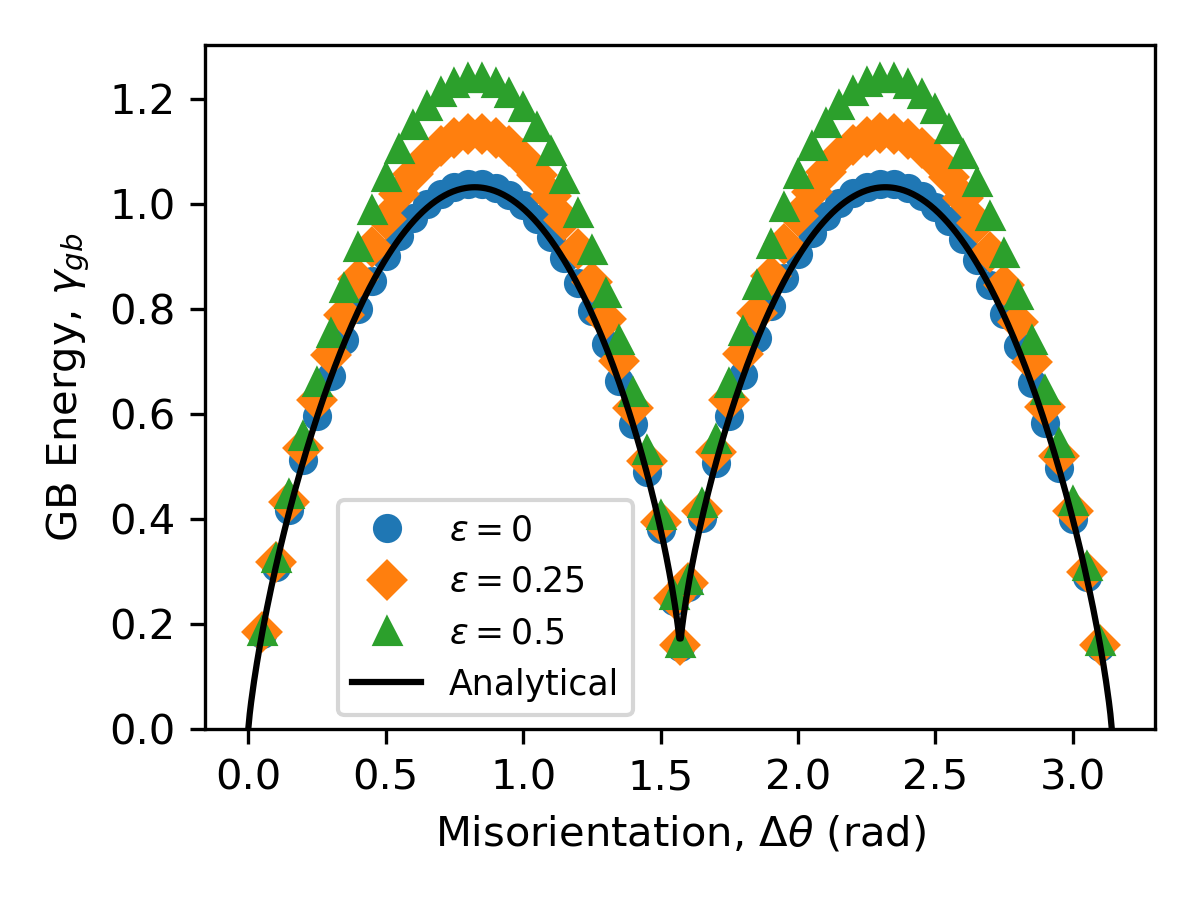}
    \caption{Grain boundary energy as a function of misorientation calculated from simulations of planar grain boundaries in 1D, compared to the input analytical function for $\gamma_{GB}(\Delta\theta)$ (Eq. (\ref{eq_simulation_gamma})).}
    \label{figure_pf_energy}
\end{figure}

Given the equilibrium profiles, the GB energy can be numerically calculated and compared to the value expected based on Eq. (\ref{eq_simulation_gamma}). Figure \ref{figure_pf_energy} shows the comparison between the simulated GB energy and Eq. (\ref{eq_simulation_gamma}) for different values of $\epsilon$. The derivation of Eq. \ref{eq_simulation_gamma} involves inverting the relationship between $\gamma_{GB}$ and $\Delta\theta$, which cannot be done analytically for $\epsilon\neq0$. Therefore, the coefficient $s$ based on Eq. \ref{eq_kwc_s_deltatheta} gives exactly the embedded $\gamma_{GB}$ only for simulations with $\epsilon=0$. As $\epsilon$ increases from zero, the GB energy in the simulations exceeds that specified by the input GB energy function. Because the KWC model requires nonzero $\epsilon$ for GBs to be mobile, the parameterization of GB energy in the present model still requires some calibration based on the selected $\epsilon$ to ensure that the desired $\gamma_{GB}(\Delta\theta)$ is accurately reproduced. Interestingly, $\gamma_{GB}(\Delta\theta)$ for $\epsilon>0$ is well-approximated with a scalar multiple of the $\epsilon=0$ GB energy curve, which can be useful for accelerating the calibration process. 

The chosen piecewise GB energy function, Eq. \ref{eq_simulation_gamma}, has a cusp exactly at $\Delta\theta=\pi/2$, which is reproduced in the numeric computations of the 1D profiles without additional regularization or approximations. The presence of the cusp in the GB energy does not cause instability, since the cusp enters through the dependence on nonlocal misorientation, and the variation of $\theta$ with the nonlocal misorientation is zero. The ability to capture sharp cusps in GB energy expands the applicability of the present KWC model to a wider variety of systems. In fact, as long as $\gamma_{GB}$ goes to zero as $\Delta\theta=0$, nearly-isotropic GB energy functions can be simulated; in theory, this allows the model to emulate the isotropic ``soap-froth'' problem for benchmarking comparisons with other phase-field approaches to simulating grain growth.

\subsection{Circular grain boundaries}

Next, we apply the model to a bicrystal composed of a shrinking circular grain embedded in an infinitely large matrix grain. This is a time-dependent simulation that utilizes the GB mobility as a function of the misorientation and the model parameters, as discussed in section \ref{planar_GB}. In the limit of $\epsilon\rightarrow0$, the GB mobility $\mathcal{M}$ within the KWC model was obtained by asymptotic analysis \cite{KWC2003}.

\begin{figure}[hbt]
    \centering
    \includegraphics[width=0.5\linewidth]{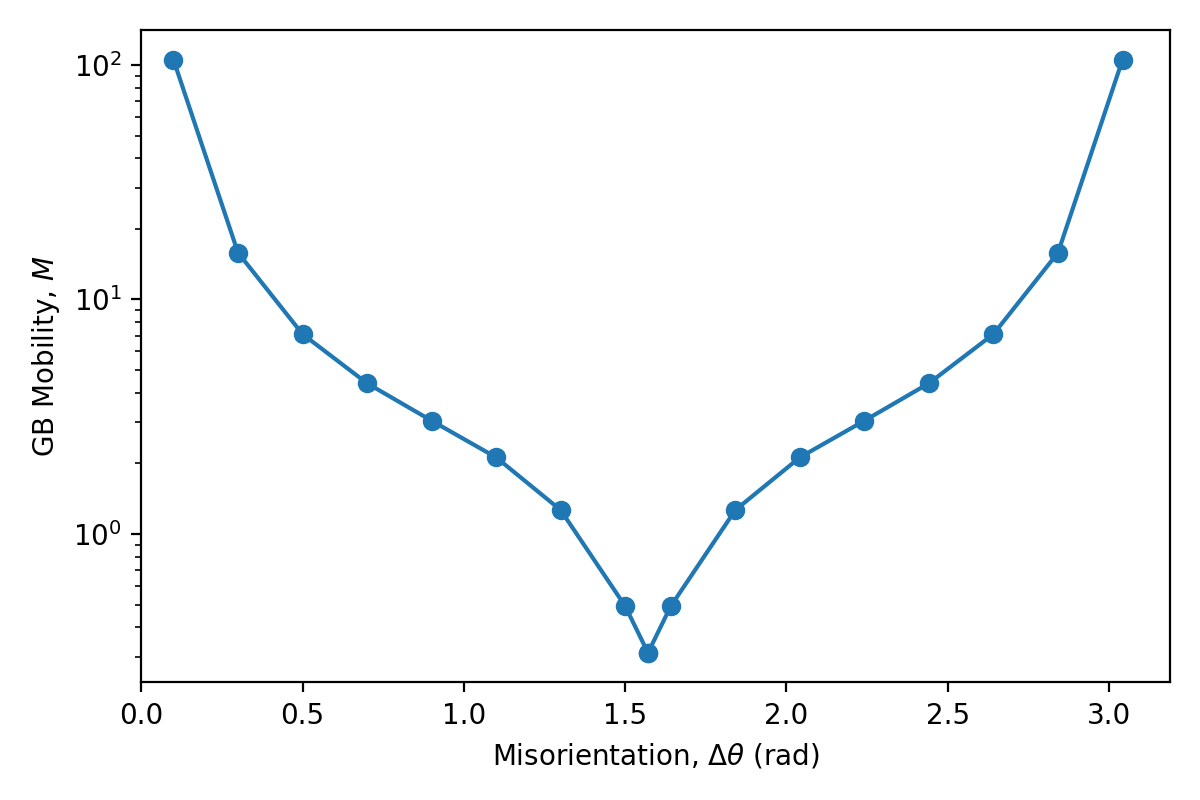}
    \caption{Grain boundary mobility as a function of misorientation, calculated from simulations of shrinking circular grains embedded in a matrix. Each point corresponds to a simulation; the line is to guide the eye.}
    \label{figure_gb_mobility}
\end{figure}

We compute the GB mobility in our current model numerically by performing simulations of circular grains embedded in larger single crystal domains. The rate of change in the area $A$ of the shrinking circular grain is related to the GB mobility by $dA/dt=-2\pi\gamma_{GB}\mathcal{M}$. The circular grain area is determined by normalizing $\theta$ by the misorientation, so that the inner grain elements are $1$ and the matrix grain elements are $0$; then, the sum of the normalized $\theta$ multiplied by $\Delta x^2$ gives the inner grain area. The GB mobility decreases with misorientation, diverging to infinity as $\Delta\theta\rightarrow0$ (Figure \ref{figure_gb_mobility}). 

The GB mobility decreases with misorientation, with a similar dependence on $\Delta\theta$ as $s(\Delta\theta)$ (compare Figure \ref{figure_gb_mobility} to \ref{figure_s_gamma}b). This resembles the mobility of a low-angle planar grain boundary moving via translation of the structure units forming the boundary \cite{Karma2012}. The situation is more complex in curved boundaries, as migration of the boundary requires either the generation or annihilation of structural units or rotation of the grains \cite{Qiu2026}. For our model, we posit that the dependence of GB mobility on misorientation can be tuned by multiplicative composition of the $\theta$ mobility coefficient, $M_\theta$, with an appropriate function of $\nabla\theta$ and/or $\Delta\theta$. This approach was demonstrated in previous work, where the $\theta$ mobility was made a function of $\phi$ and $\nabla\theta$ \cite{Staublin2022}. The details of adjusting the GB mobility in the present model are left for future work.

A more subtle feature of the KWC model is grain rotation, which can be reduced (but not completely eliminated) by adding the function $P(\epsilon|\nabla\theta|)$ to the $\theta$ mobility. What is important to emphasize, however, is that rotation can be controlled through this function, which can be used to dramatically slow evolution of the $\theta$ field in regions where $|\nabla\theta|$ is below a small value. With the present choice for $\mu$ and $\beta$, this value is approximately $|\nabla\theta|\approx0.01$. Because this value is not exactly zero, regions with small gradients in $\theta$ persist outside diffuse GBs. These small regions can propagate from moving GBs, causing slow long-range grain rotations. The magnitude of the persistent small gradient does not depend on the macroscopic GB misorientation, so small-angle GBs are affected to a greater degree compared to high-angle GBs. With the parameters used in this work, migrating GBs with misorientation below approximately 5$^\circ$ tend to adopt linear profiles with constant $|\nabla\theta|\approx0.1$ spread over a wide region.

\section{3D extension of the new model}
\label{sec_3D}

In the previous sections, the proposed model was applied to 1D cases of a planar or circular GB. Here, we outline a possible extension of the model of 3D systems.

Orientations in three dimensions have three degrees of freedom, necessitating the use of a vector field to represent the orientation field. There are a number of possible representations for 3D rotations, including Euler angles, rotation matrices, axis-angle pairs, and Rodrigues vectors \cite{MorawiecBook,Rowenhorst2015}. For computational efficiency, unit quaternions are ideal for extending the present model to 3D. A unit quaternion can be represented as a four-dimensional unit vector and describes a rotation of $\psi$ degrees about an axis $\hat{e}$:
\begin{equation}
  q_i =
  \begin{cases}
    \cos(\psi/2) & i = 0 \\
	\hat{e}_i\sin(\psi/2) & i = 1,2,3 \\
  \end{cases}
\end{equation}
A constraint must be introduced such that the norm of the quaternions remains unity.

The gradient in orientation can be replaced by the sum of squares of the quaternion components, which maintains rotational invariance. More complex forms for local misorientation can be derived that relate to the local misorientation axis and angle \cite{Admal2019, PhilThesis}; however, these forms are not necessary when introducing nonlocal misorientation. The misorientation between two orientations on either side of the grain ($-$ and $+$) represented by quaternions is simply $\mathbf{q_{-}}\mathbf{q_{+}}^{-1}$, where $\mathbf{q}^{-1}$ is the inverse of $\mathbf{q}$. Due to crystal symmetry, a given (mis)orientation can be represented by multiple quaternions. Care must be taken to ensure (mis)orientations are mapped to a fundamental zone, similarly to the mapping of angle differences onto a single period in the two-dimensional model \cite{PhilThesis}. Misorientation-dependent coefficients can then be inserted into the free energy functional:
\begin{equation}
    F[\phi,\mathbf{q}] = \int_\Omega \left[V(\phi) + \frac{\alpha^2}{2}|\nabla\phi|^2 +s(\mathbf{q_{-}}\mathbf{q_{+}}^{-1})g(\phi)\sum_{i=0}^{3}|\nabla q_i|+\frac{\epsilon^2 s(\mathbf{q_{-}}\mathbf{q_{+}}^{-1})^2}{2}h(\phi)\sum_{i=0}^{3}|\nabla q_i|^2 \right]dV.
\end{equation}
The coefficient $s$ becomes a function of 3D misorientation, with domain $SO(3)$. The symmetrized hyperspherical harmonics described by Mason and Schuh \cite{Mason2008} can be employed to construct functions on $SO(3)$ with the appropriate symmetries.

Evaluation of the nonlocal misorientation requires the grain boundary normal vector, which was expressed as $\hat{n}=\nabla\theta/|\nabla\theta|$ in 2D. In the 3D model, the order parameter can be used to compute the normal vector: $\hat{n}=\nabla\phi/|\nabla\phi|$. Alternatively, rotationally-invariant expressions involving the 3D orientation field can be derived for the local inclination vector \cite{Admal2019}. The 2D bilinear interpolation of the orientation field must be replaced by a spherical interpolation of quaternions. 

\section{Conclusions}
\label{sec_conclusions}

We have presented a proof that orientation-field models for grain coarsening (including the Kobayashi-Warren-Carter and Henry-Mellenthin-Plapp models) cannot simulate the GB free energy decreasing with misorientation, limiting their applicability to simulating polycrystalline materials with energy cusps at special boundaries. We demonstrate a method to overcome this limitation by introducing a misorientation dependence of the orientation coupling coefficient, where the misorientation is determined by a line integral through the diffuse boundary width. This measure of misorientation requires evaluating the orientation on either side of the diffuse boundary, which we achieve (inspired by the work of Han and van de Walle) by interpolating the orientation field at a constant distance along both directions of the GB normal vector. We apply this approach to the KWC model and demonstrate the capacity of the modified model to embed GB free energies as arbitrary functions of misorientation. The modified model allows for sharp cusps in the GB free energy as a function of misorientation, expanding the applicability of the KWC model to grain growth in polycrystalline materials with non-monotonic GB free energy. Because the form of the free energy function corresponds closely to the KWC model, many ``tricks'' for performing KWC simulations are readily adapted to the present work, including the suppression of grain rotation and handling of topological defects. Finally, we propose a straightforward 3D extension of the model using quaternions to represent the orientation field and hyperspherical harmonic functions to embed the misorientation-dependence of the GB free energy. Future model improvements could include alternate methods for calculating nonlocal misorientation (such as by image segmentation), incorporation of inclination dependence, and parallelization of the numerical method used to integrate the evolution equations.

\textbf{\large{Acknowledgments}}

This research was supported by the National Institute for Standards and Technology through  the Center for Hierarchical Materials Design at Northwestern University contract \newline [70NANB19H00]. This research was supported in part through the computational resources and staff contributions provided for the Quest high performance computing facility at Northwestern University which is jointly supported by the Office of the Provost, the Office for Research, and Northwestern University Information Technology.

\clearpage
\phantomsection
\bibliographystyle{unsrt}
\interlinepenalty=1000
\nocite{*}
\bibliography{references}

\end{document}